\documentclass[11pt]{article}
\usepackage{amsmath} 
\usepackage{amssymb}
\usepackage{verbatim} 
\usepackage{bm} 
\usepackage[colorlinks=true,citecolor=blue,linkcolor=black]{hyperref}
\usepackage{authblk} 

\usepackage{sectsty} 
\sectionfont{\large}
\subsectionfont{\normalsize}
\subsubsectionfont{\normalsize}
\numberwithin{equation}{section} 

\def\k{{\bf k}}

\def\x{{\bf x}}

\def\vs{v_{\rm s}}
\def\grad{{\bm\nabla}}
\def\coeff#1#2{{\textstyle {\frac {#1}{#2}}}}
\def\GG{{\cal G}}

\newcommand{\nc}{\newcommand}
\nc{\al}{\alpha}
\nc{\ga}{\gamma}
\nc{\de}{\delta}
\nc{\ep}{\epsilon}
\nc{\ze}{\zeta}
\nc{\et}{\eta}

\nc{\ka}{\kappa}
\nc{\la}{\lambda}
\nc{\rh}{\rho}
\nc{\si}{\sigma}
\nc{\ta}{\tau}
\nc{\up}{\upsilon}
\nc{\ph}{\phi}
\nc{\ch}{\chi}
\nc{\ps}{\psi}
\nc{\om}{\omega}
\nc{\Ga}{\Gamma}
\nc{\De}{\Delta}
\nc{\La}{\Lambda}
\nc{\Si}{\Sigma}
\nc{\Up}{\Upsilon}
\nc{\Ph}{\Phi}
\nc{\Ps}{\Psi}
\nc{\Om}{\Omega}
\nc{\ptl}{\partial}
\nc{\del}{\nabla}
\nc{\be}{\begin{equation}}
\nc{\ee}{\end{equation}}
\nc{\ba}{\begin{eqnarray}}
\nc{\ea}{\end{eqnarray}}
\nc{\bi}{\bibitem}

\title{ On thermal fluctuations and the generating functional in relativistic hydrodynamics
      }

\author{Michael Harder, Pavel Kovtun, and Adam Ritz}
\affil{
    Department of Physics and Astronomy, University of Victoria, Victoria, BC, V8P 5C2, Canada \vspace{-4ex}}

\date{ (February 2015)}

\begin{document}

\maketitle

\begin{abstract}
\noindent
We discuss a real-time generating functional for correlation functions in dissipative relativistic hydrodynamics which takes into account thermal fluctuations of the hydrodynamic variables. Starting from the known form of these correlation functions in the linearized regime, we integrate to find a generating functional which we can interpret within the CTP formalism, provided the space-time and internal global symmetries are realized in a specific manner in the $(r,a)$ sectors. We then verify that this symmetry realization, when implemented in an effective action for hydrodynamic fields in the $(r,a)$ basis, leads to a consistent derivative expansion for the constitutive relations at the nonlinear level, modulo constraints associated with the existence of an equilibrium state.
\end{abstract}

\section{Introduction}
Hydrodynamics is a low-energy effective description of many physical systems in local thermal equilibrium. Relativistic hydrodynamics for normal fluids is a set of partial differential equations expressing the conservation of the energy-momentum tensor and other conserved currents, such as the baryon number current. Such a description is classical, rather than field-theoretic, and one can enquire about the low-energy effective field theory corresponding to the hydrodynamic regime, and how it is related to the classical hydrodynamic equations. A fundamental object in field theory is the generating functional $W[A,g]$ where $A_\mu$ and $g_{\mu\nu}=\eta_{\mu\nu}+h_{\mu\nu}$ are the external sources (gauge field and the metric). The variations of $W[A,g]$ with respect to the sources give rise to hydrodynamic correlation functions, i.e. to correlation functions of the energy-momentum tensor $T^{\mu\nu}$ and the current $J^\mu$ in the limit of small frequency and momentum. 

The motivation for finding such a $W[A,g]$ is the following. There is a plethora of $n$-point real-time response functions, differing by time ordering and symmetrization of the corresponding operators. These correlation functions can be conveniently classified in the closed time path (CTP) formalism by labeling the operators according to the two parts of the time contour~\cite{Chou:1984es, Wang:1998wg}. The response functions computed in classical hydrodynamics by varying the on-shell $T_{\rm cl}^{\mu\nu}$ and $J_{\rm cl}^\mu$ with respect to the sources are the fully retarded functions, or $raa...a$ functions, in the notation of Ref.~\cite{Wang:1998wg}. While for two-point functions, all response functions can be reconstructed from the $ra$ function by the fluctuation-dissipation theorem, the same is not true for general $n$-point functions~\cite{Wang:1998wg}. In classical hydrodynamics, the fluctuation-dissipation relations are not contained in the hydrodynamic equations, and have to be imposed by hand.

Second, classical hydrodynamics misses information about thermal fluctuations of the hydrodynamic degrees of freedom themselves. Such real-time thermal fluctuations will lead to effects (long-time tails and momentum-space non-analyticities in two-point functions) which can not be captured by classical hydrodynamic equations~\cite{DeSchepper19741, Pomeau197563}. Relativistic fluids are not immune to such effects, though the fluctuation corrections become suppressed in the large-$N$ limit~\cite{Kovtun:2003vj, Kovtun:2011np}. A full real-time generating functional must capture both the fluctuation-dissipation theorem, and the effects of thermal fluctuations. 

One way to arrive at a hydrodynamic effective action is to modify the hydrodynamic constitutive relations by introducing random stresses and currents whose correlation properties are chosen so that the fluctuation-dissipation theorem is satisfied in equilibrium~\cite{LL9}. Such a construction is phenomenological: it takes the hydrodynamic equations as given, while the random stresses and currents become extra dynamical degrees of freedom, to be integrated over in the hydrodynamic path integral. It is not immediately clear how to proceed with the systematic derivative expansion and the coupling to external sources in this formalism. See e.g.~\cite{Kapusta:2011gt, Kovtun:2014hpa} for recent discussions in the context of relativistic hydrodynamics. From a field-theoretic perspective, it would be more natural to implement the hydrodynamic equations, the derivative expansion and the coupling to external sources at the level of the effective action which respects the relevant symmetries of the microscopic physical system.

The present paper is a step in this direction. We start with the bottom-up approach in Section~\ref{sec:bottom-up}, asking a simple question: what is the generating functional that gives rise to the known hydrodynamic two-point functions of linearized hydrodynamics in equilibrium? Incorporating the appropriate background sources, a structure emerges that is consistent with expectations from the CTP formalism, expressed in the so-called $(r,a)$ basis.
Going beyond linearized hydrodynamics requires understanding the symmetries of the effective action. In the CTP formalism, the set of symmetries is doubled (call these symmetries $\GG_1$ and $\GG_2$), corresponding to the two branches of the time contour. The classical hydrodynamic equations on the other hand manifest only one symmetry, the diagonal (physical) $\GG_r$. We discuss the symmetries and derive the relevant Ward identities in Section~\ref{sec:top-down}, which generalize the classical hydrodynamic conservation equations. These classical equations have the schematic form of conservation laws
\be
 {\cal D}\cdot {\cal J}_r=0,
\ee
where ${\cal J}_r= J^\mu, T^{\mu\nu},$ etc. and we have ignored possible explicit symmetry breaking terms that may be present on the right hand side.
Based on the results of the bottom-up analysis, we further argue that the hydrodynamic effective action can be built from the degrees of freedom that arise in a low energy nonlinear symmetry realization, analogous to the spontaneous breaking of $\GG_1 \times \GG_2 \rightarrow \GG_r$. The degrees of freedom of the effective theory thus include modes analogous to the extra Goldstone modes arising from symmetry breaking. The effective action has the following schematic expansion in terms of the $a$-sector (fluctuation) fields $\varphi_a$,
\be
 S_{\rm eff} = {\cal I}_r + {\cal J}_r \cdot {\cal D}\varphi_a + {\cal K}_r \cdot ({\cal D} \varphi_a)^2 + \dots.
\ee
Here ${\cal I}_r$, ${\cal J}_r$ and ${\cal K}_r$ are functionals of the $r$-sector (physical) fields, with indices suppressed. ${\cal D}\varphi_a$ denotes the appropriate covariant derivative of $\varphi_a$, which is manifestly invariant under the hidden symmetries $\GG_a$, orthogonal to $\GG_r$. The term linear in ${\cal D} \varphi_a$, on varying with respect to $\varphi_a$, enforces the classical equations of motion ${\cal D}\cdot {\cal J}_r=0$, a generic feature of the CTP action. The term quadratic in ${\cal D} \varphi_a$ provides the additional structure necessary to satisfy the fluctuation-dissipation theorem, and can be interpreted in terms of supplying fluctuations. It is apparent that the derivative coupling of $\varphi_a$ is consistent with the dependence expected for Goldstone modes associated with a nonlinear symmetry realization. We test this expectation at the nonlinear level
by writing the general CTP effective action in a low energy derivative expansion for both the classical hydrodynamic fields, and for the fluctuation modes $\varphi_a$. In Section~\ref{sec:derivative-expansion} we implement the expansion to first order and identify all the expected transport coefficients of first-order hydrodynamics in terms of the parameters of the effective action, modulo certain constraints associated with the existence of an equilibrium state. We conclude in Section~\ref{sec:discussion} with a list of open questions that need to be resolved in order to have a complete picture of the hydrodynamic generating functional.

\section{Bottom-up approach}
\label{sec:bottom-up}
\subsection{Diffusive mode}
We start with linearized hydrodynamics, and consider the simplest hydrodynamic process which is diffusion. It is described by the diffusion equation
\begin{equation}
\label{eq:diff-eq}
  \partial_t n - D \grad^2 n = 0\,,
\end{equation}
where $n$ is the charge density fluctuation, $D$ is the diffusion constant, and $\grad^2\equiv\partial_i \partial^i$. The linear response theory gives the following two-point functions of the charge density $n(t,\x)$ in thermal equilibrium:
\begin{equation}
\label{eq:diff-corr}
   G_{ra}(\omega,\k) = \frac{D\chi\,\k^2}{i\omega - D\k^2}\,,\ \ \ \
   G_{ar}(\omega,\k) = \frac{-D\chi\,\k^2}{i\omega + D\k^2}\,,\ \ \ \ 
   G_{rr}(\omega,\k) = \frac{-4i\, T D \chi\,\k^2}{\omega^2 + (D\k^2)^2}\,.
\end{equation}
The first one is retarded, the second one is advanced, the third one is $(-i)$ times the anti-commutator, and $G_{aa}$ is identically zero. Here $T$ is the equilibrium temperature, and $\chi \equiv (\partial n/\partial\mu)_{\mu=0}$ is the static charge susceptibility. What is the effective action which gives rise to these correlation functions? 

In relativistic hydrodynamics, the diffusion equation emerges from the current conservation equation $\partial_\mu J_{\rm cl}^\mu = 0$ in the Landau-Lifshitz frame~\cite{LL6}, linearized in small fluctuations close to the equilibrium state at zero chemical potential. We can couple the system to an external gauge field $A_\mu$, which gives rise to $J_{\rm cl}^\mu[A]$. The variation of the hydrodynamic on-shell current with respect to the source $A$ can give rise to $G_{ra}$ and $G_{ar}$, but {\em not} to $G_{rr}$. This is because the linearized current conservation equation in the presence of the source has in it $D$ and $\sigma=D\chi$, but not $T$. In order to get $G_{rr}$, hydrodynamic equations coupled to sources are not enough, and one has to use extra information, namely the fluctuation-dissipation theorem. The effective action must incorporate both the hydrodynamic equations with sources, and the fluctuation-dissipation theorem.

It is intuitively clear that there is no local path integral action $S[\phi_r,\phi_a]$ quadratic in the fields, which would give $G_{ra} = -i\langle \phi_r \phi_a \rangle$, $G_{ar} = -i\langle \phi_a \phi_r \rangle$, $G_{rr} = -2i\langle \phi_r \phi_r \rangle$ with the response functions~(\ref{eq:diff-corr}). This is because the response functions (\ref{eq:diff-corr}) have $\k^2$ in the numerator, hence the corresponding action will have $\k^2$ in the denominator, which is not local in space. For a quadratic action of the form
\begin{equation}
\label{eq:S-quadratic}
  S = \frac12 \int_{\omega,\k} \phi^{\alpha\ *}_{\omega,\k}\, P_{\alpha\beta}(\omega,\k)\, \phi^\beta_{\omega,\k}
\end{equation}
in the path integral, the matrix
\begin{equation}
\label{eq:P-diff}
  P_{\alpha\beta} = 
  \frac{1}{D \chi \k^2}
  \begin{pmatrix}
   0 & -i\omega - D\k^2 \\
   i\omega-D\k^2 & 2iT
  \end{pmatrix}
\end{equation}
gives the correct response functions~(\ref{eq:diff-corr}).
Here, the upper left element is $rr$, upper right is $ra$, bottom left is $ar$, and bottom right is $aa$. In particular, 
$$
  \langle \phi_\alpha \phi_\beta \rangle = i(P^{-1})_{\alpha\beta}\,,
$$
where the indices run over $r$, $a$. The action (\ref{eq:S-quadratic}), (\ref{eq:P-diff}) is not real: it is complex, but in such a way that the functional integral with the weight $e^{iS}$ converges. Clearly, the matrix (\ref{eq:P-diff}) is not analytic in $\k$, and the quadratic action $S[\phi_r,\phi_a]$ is not local in space. 

The action can be made local by introducing auxiliary fields. Let us define a new field $\varphi_a$ as $\phi_a = D\chi\grad^2 \varphi_a$. Consider the following action
\begin{equation}
\label{eq:S-diff}
  S[\phi_r,\phi_a,\varphi_a,\lambda] = \int\!\!dt\,d^dx \Big[
  \varphi_a (\partial_t - D\grad^2) \phi_r - i \varphi_a T \phi_a + \lambda (\phi_a - D\chi\grad^2 \varphi_a)
  \Big]\,,
\end{equation}
where the auxiliary field $\lambda$ is used to impose the constraint which defines $\varphi_a$. 
We can further define the generating functional as
$$
  Z[a_r,a_a] = e^{iW[a_r,a_a]} = 
  \int\!\!D\phi_r D\phi_a D\varphi_a D\lambda\, 
  e^{iS+i\!\int\! dt\, d^dx (a_a \phi_r + a_r \phi_a)}\,,
$$
where the effective action is given by Eq.~(\ref{eq:S-diff}), and $a_r$, $a_a$ are external sources. The physical meaning of $\phi_r$ is the density fluctuation field $n$. By construction, this generating functional reproduces~(\ref{eq:diff-corr}). Integrating out $\lambda$ and $\phi_a$ leaves
\begin{eqnarray}
\label{eq:ZZ}
 Z[a_r,a_a] = \int\!\!D\phi_r D\varphi_a \, 
  \exp \Big[ i\!\!\int_{t,\x}\!  
  \big[
  \varphi_a \big(\partial_t \phi_r - D\grad^2 \phi_r + D\chi\grad^2 a_r\big) \nonumber\\
  -i \varphi_a \, T D\chi \grad^2 \varphi_a + a_a \phi_r \big]  \Big].
\end{eqnarray}
The source $a_r$ now appears as a correction to the equation of motion of $\phi_r$. This is precisely how the source term should appear, based on the full hydrodynamic description. Indeed, in relativistic hydrodynamics, the diffusion equation (\ref{eq:diff-eq}) arises from current conservation. To first order in the derivative expansion, the constitutive relation for the current in the Landau-Lifshitz frame is
\begin{equation}
\label{eq:JJ}
  J_{\rm cl}^\mu = n u^\mu  - \sigma T \Delta^{\mu\lambda} \partial_\lambda(\mu/T)
  + \sigma \Delta^{\mu\lambda} E_\lambda\,,
\end{equation}
where $u^\mu$ is the fluid velocity satisfying $u^\mu u_\mu =-1$, $T$ is the local temperature, $\mu$ is the chemical potential, and $\Delta^{\mu\nu} = \eta^{\mu\nu}+u^\mu u^\nu$. The electric field is $E_\mu = F_{\mu\nu} u^\nu$, where $F_{\mu\nu}$ is the field strength. Let us turn on $A_0$ only. For linearized fluctuations around the equilibrium state with $u^\mu = (1,{\bf 0})$, $T={\rm const}$, and $\mu=0$, we have $J_{\rm cl}^0 = n$, $J_{\rm cl}^i = -\sigma \partial^i \mu + \sigma \partial^i A_0$. Charge density and chemical potential fluctuations are related by $n=\chi\mu$, and the electrical conductivity is $\sigma=D\chi$. Current conservation $\partial_\mu J_{\rm cl}^\mu=0$ now gives
$$
  \partial_t n - D\grad^2 n + D\chi \grad^2 \!A_0 = 0\,,
$$
modifying the diffusion equation by the source term proportional to $\grad^2 A_0$. Thus, by comparing with (\ref{eq:ZZ}), we identify $a_r$ with $A_0^r$. The hydrodynamic equation coupled to the source contains both $D$ and $\chi$, but not $T$. The dependence on temperature comes from the kinetic term for $\varphi_a$ in the effective action (\ref{eq:ZZ}).

One can rewrite the kinetic term for $\varphi_a$ as
\begin{equation}
\label{eq:psitheta}
  e^{TD\chi\int\! \varphi_a \grad^2 \varphi_a} = \int\!\!Dr\; 
  e^{\frac{1}{4TD\chi} \int\! r_i r_i }
  e^{i\!\int\! \varphi_a \partial_k r_k}\,.
\end{equation}
This makes the action linear in $\varphi_a$, enforcing the equation of motion for $\phi_r$ with the Gaussian noise $r_i$ in the right-hand side. Integrating out $\varphi_a$ gives
$$
  Z[a_r, a_a] = \langle e^{i\int\! a_a\, n[a_r,r]}\rangle_{r}\,,
$$
where $n[a_r,r]$ in the exponent is a solution to $\partial_t n - D\grad^2 n + D\chi\grad^2 a_r = -\partial_k r_k$, for a given source $a_r(t,\x)$ and noise profile $r_i(t,\x)$, and the average is over the Gaussian noise~$r_i$. This is the standard relation between stochastic equations and path integrals. 

One can write down a covariant generalization of the generating functional~(\ref{eq:ZZ}),
\begin{equation}
\label{eq:ZZA}
  Z[A_r,A_a] = \int\!\! D\phi_r D\varphi_a\; e^{iS[\phi_r,\varphi_a, A_r, A_a]}\,,
\end{equation}
where 
\begin{equation}
\label{eq:S-lin-A}
  S =  \int\!\!dt\,d^dx \left( J_{\rm cl}^\mu[\phi_r, A_r] D_\mu \varphi_a 
   +i T\sigma \Delta^{\mu\nu} D_\mu \varphi_a\, D_\nu \varphi_a 
   \right)\,,
\end{equation}
and $D_\mu \varphi_a \equiv \partial_\mu \varphi_a + A_\mu^a$. The gauge field $A_r$ is the physical gauge field, while $A_a$ is only used as a tool to access correlation functions, and needs to be set to zero at the end of the calculation. The effective action is invariant with respect to both $r$-type and $a$-type gauge transformations. The conserved current obtained by varying the effective action with respect to $A_a$ is the classical hydrodynamic current $J_{\rm cl}^\mu$ plus the fluctuation correction,
\begin{equation}
  J^\mu = J_{\rm cl}^\mu + 2iT\sigma \Delta^{\mu\nu} D_\nu \varphi_a\,.
\end{equation}

\subsection{Shear and sound modes}
Let us now look at the hydrodynamic shear and sound modes. It will be easiest to work with an uncharged fluid. In the Landau-Lifshitz frame, the energy density fluctuation $\epsilon= T^{00}-\bar\epsilon$ (with $\bar\epsilon$ the equilibrium value) and the momentum density fluctuation $\pi^i=T^{0i}$ obey
\begin{eqnarray*}
  && \partial_t \epsilon + \partial_k \pi_k = 0\,,\\
  && \partial_t \pi_i + \vs^2 \partial_i \epsilon - M_{ij}\pi_j = 0\,,
\end{eqnarray*}
where $\vs^2=\partial\bar p/\partial\bar\epsilon$ is the speed of sound squared, $M_{ij} = \gamma_\eta(\grad^2\delta_{ij} - \partial_i \partial_j) + \gamma_s\partial_i \partial_j$, and the damping coefficients are $\gamma_\eta=\eta/(\bar\epsilon+\bar p)$, $\gamma_\zeta=\zeta/(\bar\epsilon+\bar p)$, $\gamma_s=\gamma_\zeta+\frac{2d-2}{d}\gamma_\eta$. The hydrodynamic retarded functions are
\begin{subequations}
\label{eq:GG-shear-sound}
\begin{equation}
\label{eq:GRpiipij}
  G^{\,ra}_{\pi_i \pi_j}(\omega,\k) = 
  \left(\delta_{ij} - \frac{k_i k_j}{\k^2}\right)
  \left[ \frac{\bar w \gamma_\eta \k^2}{i\omega-\gamma_\eta \k^2} + \bar\epsilon \right]
  +
  \frac{k_i k_j}{\k^2}\,
  \left[\frac{\bar w \omega^2}
  {\omega^2-\k^2\vs^2+i\omega\gamma_s \k^2} -\bar p \right]
  \,,
\end{equation}
\begin{eqnarray}
  &&  G^{\,ra}_{\epsilon \pi_i}(\omega,\k) = G^{\,ra}_{\pi_i \epsilon}(\omega,\k) =
      \frac{\bar w\, \omega k_i}{\omega^2-\k^2\vs^2+i\omega\gamma_s \k^2}\,,
\label{eq:GRepii}\\[5pt]
  &&  G^{\,ra}_{\epsilon\epsilon}(\omega,\k) = 
      \frac{\bar w\, \k^2}{\omega^2-\k^2\vs^2+i\omega\gamma_s \k^2} - \bar\epsilon\,,
\label{eq:GRee}
\end{eqnarray}
\end{subequations}
where $\bar w = \bar\epsilon+\bar p$. These functions are obtained by first solving the hydrodynamic equations in the external metric, and then by varying the resulting solution for $\sqrt{-g}\,T^{\mu\nu}$ with respect to the metric, see for example~\cite{Kovtun:2012rj}. Note that $G^{\,ra}_{\pi_i \pi_j}(\omega,\k)$ is analytic as $\k\to0$. All $aa$ functions vanish identically, while the equilibrium $rr$ functions can be obtained from the fluctuation-dissipation theorem as
\begin{equation}
\label{eq:Grr2}
  G^{\,rr}_{AB} = \frac{4iT}{\omega} {\rm Im}\, G^{\,ra}_{AB}\,.
\end{equation}
We would like to find an effective action which reproduces the above response functions, as well as the corresponding $rr$ functions. Let us choose $\epsilon$ and $\pi_i$ as our variables. By analogy with the diffusive generating functional (\ref{eq:ZZ}), one can make a guess for the generating functional for the shear and sound modes:
\begin{align}
\label{eq:ZZ2}
  Z[h_r,h_a] &= \int\!\!D\epsilon D\pi_i D\varphi^a_0 D\varphi^a_i \, 
  \exp \Big[i\!\!\int_{t,\x}\! 
  \big[
     \varphi^a_i \big(\partial_t \pi_i +\vs^2\partial_i \epsilon - M_{ij} \pi_j 
     + \bar w\, \partial_t h_{0i}^r -\coeff12 \bar w\, \partial_i h_{00}^r \big) \nonumber\\
     &\quad +\varphi^a_0 \big( \partial_t\epsilon {+} \partial_k \pi^k \big)
   -i T \bar w\, \varphi^a_i M_{ij} \varphi^a_j  + \coeff12 h_{00}^a (\bar\epsilon+\epsilon+\coeff12\bar\epsilon h_{00}^r) + h_{0i}^a (\pi^i {+} \bar p h_{0i}^r) \big] \Big].
\end{align}
The fields $\epsilon,\pi_i$ here are $r$-type fields, and the auxiliary fields $\varphi^a_0$, $\varphi^a_i$ are $a$-type fields.  The only sources turned on are $h_{0\mu}$. The $h_{0\mu}^r$ sources in the equation of motion come from $\nabla_\mu T^{\mu\nu}=0$ in the Landau-Lifshitz frame. The $h_{0\mu}^a$ sources come from 
$\coeff12 \sqrt{-g_r}\, T^{\mu\nu}_r h_{\mu\nu}^a$.

An exercise with Gaussian integrals, given in Appendix A, shows that the generating functional (\ref{eq:ZZ2}) does indeed reproduce the equilibrium response functions (\ref{eq:GG-shear-sound}) of linearized hydrodynamics in an uncharged relativistic fluid. The fluctuation-dissipation theorem (\ref{eq:Grr2}) and the vanishing of all $aa$ functions automatically follow from the structure of the effective action in~(\ref{eq:ZZ2}). Just as in the example of diffusion, the generating functional can be cast into the form of a stochastic equation with Gaussian noise.

One can write down a covariant generalization of the generating functional~(\ref{eq:ZZ2}),
\begin{equation}
\label{eq:ZZh}
  Z[h_r, h_a] = \int \!\!D\phi_\mu^r\, D\varphi^a_\mu \;
  e^{iS[\phi^r,\, \varphi^a,\, h_r,\, h_a]}\,,
\end{equation}
where 
\begin{equation}
\label{eq:S-lin}
  S = \int\!\!dt\,d^dx\; \sqrt{-g_r} \left(
  \, T_{\rm cl}^{\mu\nu}[\phi_r, g_r] \,{\cal D}_{\!\mu} \varphi^a_\nu
  +iT\, {\cal D}_{\!\mu} \varphi^a_\nu\, G^{\mu\nu\alpha\beta}\, {\cal D}_{\!\alpha} \varphi^a_\beta
  \right)\,.
\end{equation}
Here ${\cal D}_{\!\mu} \varphi^a_\nu \equiv \frac12(h^a_{\mu\nu} - \nabla_{\!\mu} \varphi^a_\nu - \nabla_{\!\nu}\varphi^a_\mu)$. 
To first order in the derivative expansion, the classical energy-momentum tensor is given by the standard expression in the Landau-Lifshitz frame,
$$
  T_{\rm cl}^{\mu\nu} = \epsilon u^\mu u^\nu + p\Delta^{\mu\nu} - G^{\mu\nu\rho\sigma} \nabla_\rho u_\sigma\,,
$$
with $G^{\mu\nu\alpha\beta} = \eta (\Delta^{\mu\alpha} \Delta^{\nu\beta} + \Delta^{\mu\alpha}\Delta^{\nu\beta} - \coeff2d \Delta^{\mu\nu} \Delta^{\alpha\beta}) + \zeta \Delta^{\mu\nu}\Delta^{\alpha\beta}$, and $\Delta^{\mu\nu} = g^{\mu\nu}_r + u^\mu u^\nu$. The indices are raised using $g_r$, which is the physical metric, while $h_a$ is only used as a tool to access correlation functions, and needs to be set to zero at the end of the calculation. The effective action is invariant with respect to both $r$-type and $a$-type diffeo transformations. The latter act as position-dependent shifts of $\varphi^a_\mu$. The conserved energy-momentum tensor obtained by varying the effective action with respect to $h^a_{\mu\nu}$ is the classical hydrodynamic $T^{\mu\nu}_{\rm cl}$ plus the fluctuation correction,
\begin{equation}
\label{eq:Tfluct}
 T^{\mu\nu} = T^{\mu\nu}_{\rm cl} + 2iT G^{\mu\nu\rho\sigma} {\cal D}_\rho \varphi^a_\sigma\,.
\end{equation}
One can explicitly check that the generating functional (\ref{eq:ZZh}) gives the correct equilibrium two-point correlation functions for all components of the energy-momentum tensor in linearized relativistic hydrodynamics, and that the fluctuation-dissipation relations are satisfied. This is a non-trivial check of the validity of the effective action~(\ref{eq:S-lin}) for linearized hydrodynamic fluctuations.

The structure of the effective action~(\ref{eq:S-lin}) is easy to discern. Suppose the kinetic terms for $\varphi^a_\mu$ were not there. Then integrating over the $\varphi^a_\mu$ would impose $\nabla_{\!\mu} T_{\rm cl}^{\mu\nu}=0$ as an exact operator equation. The generating functional then becomes
$$
  Z[h_r,h_a] = 
  e^{\frac{i}{2} \int\! \sqrt{-g_r}\, T^{\mu\nu}_\textrm{on-shell}[g_r] h_{\mu\nu}^a }
$$
where $T^{\mu\nu}_\textrm{on-shell}[g_r]$ stands for $T^{\mu\nu}[T[g_r],u[g_r],g_r]$. Taking the variation with respect to $h_{\mu\nu}^a$ produces one $r$-insertion of the on-shell $T^{\mu\nu}$, and subsequent variations with respect to $h^r$ will produce $raa\ldots a$ hydrodynamic functions, obtained in the standard way by varying the on-shell energy-momentum tensor.
The kinetic terms for $\varphi^a_\mu$ are responsible for the fluctuation-dissipation theorem, and allow one to evaluate correlation functions with more than one $r$-insertion. These terms are responsible for thermal fluctuations of the hydrodynamic modes allowing them to go off-shell, and will give rise to hydrodynamic loop corrections and running of transport coefficients.

\section{Top-down approach}
\label{sec:top-down}

In order to correctly extend the generating functional $W=-i\ln Z$ beyond Eq.~(\ref{eq:ZZh}) of linear hydrodynamics, one needs to be more systematic about the underlying symmetries.
The effective action identified in Section 2 has the schematic form,
\be
 S_{\rm eff} = {\cal I}_r + {\cal J}_r \cdot {\cal D}\varphi_a + {\cal K}_r \cdot ({\cal D} \varphi_a)^2 + \dots
 \label{S1}
\ee
Here ${\cal I}_r$, ${\cal J}_r$ and ${\cal K}_r$ are functionals of the $r$-sector fields, with indices suppressed. The term ${\cal I}_r$, which is independent of $\varphi_a$, was not present in the earlier linearized discussion, but we include it here for completeness of the expansion and will return to it below in the context of the path integral measure. The term linear in ${\cal D} \varphi_a$, on varying with respect to $\varphi_a$, enforces the classical equations of motion ${\cal D}\cdot {\cal J}_r=0$. The term quadratic in ${\cal D} \varphi_a$ supplies fluctuation corrections, and provides the additional structure necessary to satisfy the fluctuation-dissipation theorem.

\subsection{Non-equilibrium CTP contour}

We identify the expansion ({\ref{S1}) as one that appears naturally within the nonequilibrium Schwinger-Keldysh CTP formalism, involving a doubled set of fields and symmetries corresponding to two time contours, see e.g.~\cite{Chou:1984es, Wang:1998wg}. 

For a quantum-mechanical system with fundamental degrees of freedom $q$,  which at time $t_0$ is characterized by the density operator~$\rho$, the CTP generating functional is given by the path integral over the fundamental fields
\begin{equation}
\label{eq:Z-CTP}
  Z_\rho[j_1, j_2] = \int\!d\tilde q_1\,d\tilde q_2\,dq_f\; \langle \tilde q_1|\rho|\tilde q_2\rangle \int\! Dq_1 Dq_2\; e^{i\int_{t_0}^{t_f} L(q_1,j_1)} e^{-i\int_{t_0}^{t_f} L(q_2,j_2)}\,,
\end{equation}
where $j_1$ and $j_2$ are the external non-dynamical sources, and the boundary conditions are $q_1(t_0)=\tilde q_1$, $q_2(t_0)=\tilde q_2$, $q_1(t_f)=q_2(t_f) = q_f$. The generating functional satisfies $Z[j_1,j_1] = 1$ as well as $Z[j_1, j_2]^* = Z[j_2, j_1]$, thanks to ${\rm tr}\rho=1$ and $\rho=\rho^\dagger$. In terms of the $(r,a)$ variables $q_r=(q_1{+}q_2)/2$, $q_a=q_1{-}q_2$, the action is 
\be
  S[q_1, j_1] - S[q_2, j_2] = \int \!q_a E(q_r, j_r) + {\cal O}(j_a, q_a^2)\,,
  \label{eq:CTP-order-a}
\ee
where $E(q_r,j_r)$ is the classical equation of motion, see e.g.~\cite{Jeon:2004dh,Jeon:2013zga}. For a thermal equilibrium state, the temperature dependence comes from the matrix element of the density operator.  A symmetry~$\GG$ of the classical action will lead to a doubled symmetry $\GG_1 \times \GG_2$ of the generating functional provided the integration measure is invariant, and the density operator $\rho$ transforms covariantly. For the latter to be true, $\GG_1 \times \GG_2$ must reduce to the diagonal $\GG_r$ at the initial time $t_0$, as there is only one (physical) symmetry characterizing the initial state.\footnote{Note that the doubled symmetry of the generating functional does not mean that the symmetries of the theory magically double. The two-source generating functional is just a means for convenient classification of correlation functions, and the time contour can be chosen to run back and forth more than once.} For local symmetries characterized by a continuous parameter~$\xi$, we have
$$
  Z_\rho[j_1, j_2] = Z_{\rho'}[j_1 + \delta_{\xi_1} j_1, j_2+\delta_{\xi_2} j_2]\,,
$$
with $\xi_1(t_0) = \xi_2(t_0)$, $\xi_1(t_f) = \xi_2(t_f)$, and $\rho'=\delta_{\xi(t_0)}\rho$.

We seek to find a similar two-source generating functional for low-energy excitations of near-equilibrium states, in which case the effective degrees of freedom are the hydrodynamic modes (in place of~$q_r$) and the corresponding auxiliary fields (in place of~$q_a$). The classical equations of motion for the $r$-type fields are the conservation laws for the energy-momentum tensor and other currents. Note that the condition $Z[j_1,j_1] = 1$, ensured in the microscopic theory by the normalization of the initial density matrix, may result in a nontrivial measure factor at ${\cal O}(a^0)$ when we consider a low energy effective action in place of the microscopic action (\ref{eq:CTP-order-a}). In this way, we observe that the $a$-sector expansion of (\ref{S1}) can naturally emerge from the CTP formalism. The doubled symmetry $\GG_1 \times \GG_2$ of the microscopic description  (with the off-diagonal $\GG_a$ broken by the initial state) needs to be realized in terms of the effective degrees of freedom. We will subsequently propose that the correct symmetry realization in the hydrodynamic effective theory is a nonlinear realization of $\GG_1 \times \GG_2$ with an explicit (linear) realization of the diagonal $\GG_r$. The auxiliary $a$-type degrees of freedom then couple in the manner expected for the corresponding Goldstone modes.

\subsection{CTP metric sources and conservation laws}

To analyze the hydrodynamic regime, we need to consider the CTP formalism in the presence of sources for the charge current and energy momentum tensor. To that end we first introduce two metric sources $g_{\mu\nu}^1$ and $g_{\mu\nu}^2$, so that the generating functional is $W[g^1, g^2]$. We demand that $W$ is invariant under two sets of diffeomorphisms: ${\cal D}_1$ which only transforms $g^1$,
and ${\cal D}_2$ which only transforms $g^2$.
We anticipate that there will be additional fields in the theory, such that at low energies we can identify physical ($r$-sector) fluctuating modes associated with the hydrodynamic degrees of freedom, e.g. the temperature, fluid velocity, etc. We will not need to specify these modes explicitly, but we necessarily assume that their dynamics is consistent with ${\cal D}_1$ and ${\cal D}_2$. As one example, we might envisage a system with an initial equilibrium state, with inverse temperature $\bar\beta_1$ and $\bar\beta_2$ in the two sectors. The corresponding time contour is depicted in Fig.~\ref{fig:contour}. 
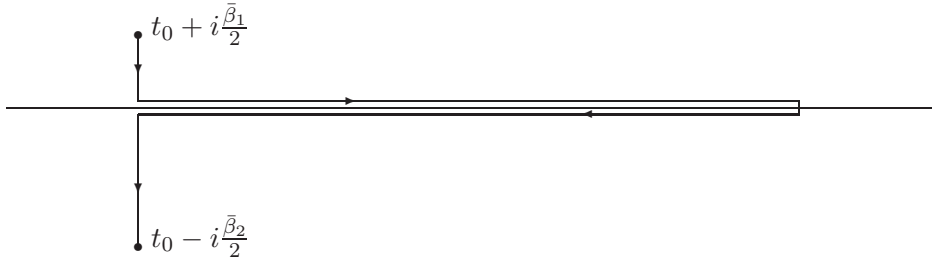
\begin{figure}
\begin{center}
\setlength{\unitlength}{2.5pt}
\begin{picture}(100,30)(0,-20)
\put(0,10){\line(0,-1){10}}
\put(0,10){\vector(0,-1){6}}
\put(0,0){\line(1,0){100}}
\put(30,0){\vector(1,0){3}}
\put(100,0){\line(0,-1){2}}
\put(100,-2){\line(-1,0){100}}
\put(70,-2){\vector(-1,0){3}}
\put(0,-2){\line(0,-1){20}}
\put(0,-2){\vector(0,-1){12}}
\put(-20,-1){\line(1,0){140}}
\put(0,10){\circle*{1}}
\put(0,-22){\circle*{1}}
\put(2,10){$t_0 + i\frac{\bar\beta_1}{2}$}
\put(2,-22){$t_0 - i\frac{\bar\beta_2}{2}$}
\end{picture}
\end{center}
\caption{A contour with two times and two $\bar\beta$'s.}
\label{fig:contour}
\end{figure}

We will be working with linear combinations of the sources which provide easy access to the retarded and symmetrized functions. We define
$$
  g^r \equiv \coeff12 \left( g^1 + g^2\right)\,,\ \ \ \ 
  g^a \equiv g^1 - g^2\,.
$$ 
The $g^r$ source corresponds to the physical metric. The $g^a$ source needs to be set to zero at the end of the calculation. Diffeo invariance of $W$ gives rise to conservations laws of the energy-momentum tensor. We define 
\begin{equation}
  \delta_g W[g^1,g^2] = \int \!\coeff12 \sqrt{-g^1}\, \langle T^{\mu\nu}_1\rangle \delta g_{\mu\nu}^1 - 
  \int \!\coeff12 \sqrt{-g^2}\, \langle T^{\mu\nu}_2\rangle \delta g_{\mu\nu}^2\,,
\end{equation}
where
\begin{align*}
 & \delta g_{\mu\nu}^1 = g_{\mu\lambda}^1 \partial_\nu\xi^\lambda_1 +
   g_{\nu\lambda}^1 \partial_\mu\xi^\lambda_1 + 
   \partial_\lambda g_{\mu\nu}^1\, \xi^\lambda_1\,,\ \ \ \ 
 & \delta g_{\mu\nu}^2 = g_{\mu\lambda}^2 \partial_\nu\xi^\lambda_2 +
   g_{\nu\lambda}^2 \partial_\mu\xi^\lambda_2 + 
   \partial_\lambda g_{\mu\nu}^2\, \xi^\lambda_2
\end{align*}
are the variations of the metric. Diffeo invariance of $W$ gives
\begin{equation}
\label{eq:dT12}
  \nabla_{\!\mu}^1 \langle T^{\mu\nu}_1\rangle = 0\,,\ \ \ \
  \nabla_{\!\mu}^2 \langle T^{\mu\nu}_2\rangle = 0\,.
\end{equation}
These conservation equations can be expressed in the $r$ and $a$ basis. To that end, we define
$$
  \sqrt{-g^r}\, \langle T^{\mu\nu}_r\rangle \equiv 
    \coeff12 \sqrt{-g^1}\, \langle T^{\mu\nu}_1\rangle + \coeff12 \sqrt{-g^2}\, \langle T^{\mu\nu}_2\rangle \,,\ \ \ \ 
  \sqrt{-g^r}\, \langle T^{\mu\nu}_a\rangle \equiv 
  \sqrt{-g^1}\, \langle T^{\mu\nu}_1\rangle - \sqrt{-g^2}\, \langle T^{\mu\nu}_2\rangle\,.
$$
The average $\langle T^{\mu\nu}_r\rangle$ is the physical stress tensor. Thus
\begin{equation}
\label{eq:deltaWra}
  \delta_g W[g^1,g^2] = \int \!\coeff12 \sqrt{-g_r}\, \langle T^{\mu\nu}_r\rangle \delta g_{\mu\nu}^a + 
  \int \!\coeff12 \sqrt{-g_r}\, \langle T^{\mu\nu}_a\rangle \delta g_{\mu\nu}^r\,.
\end{equation}
The variations $\delta g^r$ and $\delta g^a$ can be expressed in terms of $\xi_r\equiv (\xi_1{+}\xi_2)/2$ and $\xi_a\equiv \xi_1{-}\xi_2$. For the $r$-metric we have $\delta g_{\mu\nu}^r = \delta_r g_{\mu\nu}^r + \delta_a g_{\mu\nu}^r$, where
\begin{align*}
 & \delta_r g_{\mu\nu}^r = g_{\mu\lambda}^r \partial_\nu\xi^\lambda_r +
   g_{\nu\lambda}^r \partial_\mu\xi^\lambda_r + 
   \partial_\lambda g_{\mu\nu}^r\, \xi^\lambda_r\,,\ \ \ \ 
 & \delta_a g_{\mu\nu}^r = \frac14 \left( g_{\mu\lambda}^a \partial_\nu\xi^\lambda_a +
   g_{\nu\lambda}^a \partial_\mu\xi^\lambda_a + 
   \partial_\lambda g_{\mu\nu}^a\, \xi^\lambda_a \right)\,.
\end{align*}
Similarly, for the $a$-type metric we have $\delta g_{\mu\nu}^a = \delta_r g_{\mu\nu}^a + \delta_a g_{\mu\nu}^a$
where 
\begin{align*}
 & \delta_r g_{\mu\nu}^a = g_{\mu\lambda}^a \partial_\nu\xi^\lambda_r +
   g_{\nu\lambda}^a \partial_\mu\xi^\lambda_r + 
   \partial_\lambda g_{\mu\nu}^a\, \xi^\lambda_r\,,\ \ \ \ 
 & \delta_a g_{\mu\nu}^a =  g_{\mu\lambda}^r \partial_\nu\xi^\lambda_a +
   g_{\nu\lambda}^r \partial_\mu\xi^\lambda_a + 
   \partial_\lambda g_{\mu\nu}^r\, \xi^\lambda_a \,.
\end{align*}
Diffeo invariance of the generating functional (\ref{eq:deltaWra}) then gives rise to the conservation laws for $\langle T^{\mu\nu}_r\rangle$ and $\langle T^{\mu\nu}_a\rangle$. Upon setting $g^a=0$ these reduce to 
\be
\nabla_\mu \langle T^{\mu\nu}_r\rangle=0, \qquad \nabla_\mu \langle T^{\mu\nu}_a\rangle=0,
\ee 
where by $\nabla_\mu$ we denote the physical covariant derivative, evaluated with respect to $g^r$.
 
It is straightforward to incorporate external gauge fields $A_\mu^1$, $A_\mu^2$, and the corresponding combinations $A^r \equiv \coeff12(A^1+A^2)$, $A^a\equiv A^1-A^2$. 
The currents are defined by
$$
  \delta_A W = \int\!\!\sqrt{-g^1} \langle J^\mu_1 \rangle \delta A_\mu^1 - \int\!\!\sqrt{-g^2} \langle J^\mu_2 \rangle \delta A_\mu^2\,.
$$
Gauge invariance in the 1 and 2 sectors leads to 
$$
  \nabla_\mu^1 \langle J^\mu_1\rangle =0\,,\ \ \ \ 
  \nabla_\mu^2 \langle J^\mu_2\rangle = 0\,,
$$
while diffeo invariance leads to (\ref{eq:dT12}) with the usual Joule heating terms in the right-hand side. Again, the conservation laws can be expressed in the $(r,a)$ basis. We define
$$
  \sqrt{-g^r}\, \langle J^{\mu}_r\rangle \equiv 
  \coeff12 \sqrt{-g^1}\, \langle J^{\mu}_1\rangle + \coeff12 \sqrt{-g^2}\, \langle J^{\mu}_2\rangle \,,\ \ \ \ 
  \sqrt{-g^r}\, \langle J^{\mu}_a\rangle \equiv 
  \sqrt{-g^1}\, \langle J^{\mu}_1\rangle - \sqrt{-g^2}\, \langle J^{\mu}_2\rangle\,.
$$
The average $\langle J^{\mu}_r\rangle$ is the physical current.
Gauge invariance of $W$ then gives
$$
  \nabla_\mu \langle J^{\mu}_r\rangle = 0\,,\ \ \ \ 
  \nabla_\mu \langle J^{\mu}_a\rangle = 0\,.
$$
The diffeo transformation properties of the $r$ and $a$ gauge fields are $\delta A_{\mu}^r = \delta_r A_{\mu}^r + \delta_a A_{\mu}^r$, with
\begin{subequations}
\begin{align}
  \delta_r A_\mu^r = \xi^\nu_r \partial_\nu A_\mu^r + A_\nu^r \partial_\mu \xi^\nu_r\,,\ \ \ \
  \delta_a A_\mu^r = \coeff14 \left( \xi^\nu_a \partial_\nu A_\mu^a + A_\nu^a \partial_\mu \xi^\nu_a \right)\,,
\end{align}
as well as $\delta A_{\mu}^a = \delta_r A_{\mu}^a + \delta_a A_{\mu}^a$, with
\begin{align}
\label{eq:dAa}
  \delta_r A_\mu^a = \xi^\nu_r \partial_\nu A_\mu^a + A_\nu^a \partial_\mu \xi^\nu_r\,,\ \ \ \
  \delta_a A_\mu^a =  \xi^\nu_a \partial_\nu A_\mu^r + A_\nu^r \partial_\mu \xi^\nu_a \,.
\end{align}
\end{subequations}
Diffeo invariance of the generating functional then gives rise to the following conservation laws:%
\begin{subequations}
\label{eq:WI-ra}
\begin{align}
\label{eq:Tr-conserv}
  & g_{\nu\lambda}^r \nabla_\mu \langle T^{\mu\nu}_r\rangle  + \coeff14 \left( 
    g_{\nu\lambda}^a \partial_\mu \langle T^{\mu\nu}_a\rangle 
    + \Gamma_{\lambda\mu\nu}^a \langle T^{\mu\nu}_a\rangle 
    + g_{\nu\lambda}^a \Gamma^{\rho\; r}_{\rho\mu} T^{\mu\nu}_a \right) 
    = F_{\lambda\mu}^r \langle J^\mu_r \rangle + \coeff14 F_{\lambda\mu}^a \langle J^\mu_a \rangle\,,\\[5pt]
\label{eq:Ta-conserv}
  & g_{\nu\lambda}^r \nabla_\mu \langle T^{\mu\nu}_a\rangle  +  
    g_{\nu\lambda}^a \partial_\mu \langle T^{\mu\nu}_r\rangle 
    + \Gamma_{\lambda\mu\nu}^a \langle T^{\mu\nu}_r\rangle 
    + g_{\nu\lambda}^a \Gamma^{\rho\; r}_{\rho\mu} \langle T^{\mu\nu}_r\rangle 
    = F_{\lambda\mu}^a \langle J^\mu_r \rangle + F_{\lambda\mu}^r \langle J^\mu_a \rangle \,,
\end{align}
\end{subequations}
where $\Gamma_{\lambda\mu\nu} = \frac12(\partial_\mu g_{\nu\lambda} + \partial_\nu g_{\mu\lambda} - \partial_\lambda g_{\mu\nu})$. 
We have written the conservation laws in this form to avoid using the inverse of the $a$-type metric. Taking further variations with respect to the metric will give rise to Ward identities for two- and higher-point correlation functions  in the $ra$ basis.

\section{Derivative expansion for the effective action}
\label{sec:derivative-expansion}

To map the general analysis of the CTP effective action onto classical hydrodynamics, we need to understand how diffeomorphism (and/or gauge) invariance is realized in the low energy (hydrodynamic) regime. 

Let us first consider extending the straightforward realization of diffeomorphism invariance in an equilibrium state \cite{Banerjee:2012iz, Jensen:2012jh} to the hydrodynamic regime. We can characterize all near equilibrium states in terms of a timeline vector field $\beta^\mu=\bar\beta^\mu + \beta'^\mu$, where $\bar\beta^\mu$ is a timelike Killing vector characterizing the equilibrium state. If we denote the generating functional in this state as $\Gamma[g,\beta]$,\footnote{In equilibrium states, the generating functional of \cite{Jensen:2012jh} is identified here as $\Gamma[g,\bar\beta] = W[g,S] - \int\!\! \sqrt{-g}\,\bar\beta^\mu S_\mu$, where $S^\mu$ is a source for the field $\beta^\mu$, so that $ \bar\beta^\mu = \frac{1}{\sqrt{-g}}\frac{\delta W[g,S]}{\delta S_\mu}$. For equilibrium states with no source, $S_\mu = 0$, the energy-momentum tensor is conserved.} then for sufficiently well-behaved states, a derivative expansion for $\Gamma[g,\beta]$ exhibiting manifest diffeomorphism invariance can be implemented. 
As an example, consider terms up to first order in the derivative expansion. We take $\beta^\mu = u^\mu/T$, with $T=1/\sqrt{-\beta{\cdot}\beta}$, then to first order
$$
  \Gamma = \int\!d^{d+1}x\, \sqrt{-g}\, \Big( 
  p(T) + a(T) \nabla{\cdot} u + b(T) \dot T + \dots\Big)\,,
$$
where $\dot T = u^\mu\partial_\mu T$, and $p,a,b$ are arbitrary functions of~$T$. In fact, integration by parts shows that the two structures $\dot{T}$ and $\nabla \cdot u$ are not independent; the independent coefficient can be identified as the combination $(b-a')$. Varying $\Gamma$ with respect to the metric (keeping $\beta^\mu$ fixed) gives the following energy-momentum tensor: 
\begin{align}
  {T}^{\mu\nu} &= p g^{\mu\nu} + T p' u^\mu u^\nu +  (b{-}a') \underbrace{\Big( \Delta^{\mu\nu} \dot T  - T u^\mu u^\nu \nabla{\cdot} u \Big)}_{{\rm frame-dependent}},
\end{align}
where $\Delta^{\mu\nu} = g^{\mu\nu}+ u^\mu u^\nu$ is the usual transverse projector. Identifying $p(T)$ with pressure, the first two terms give the standard ideal hydrodynamics, as one can see from the thermodynamic relation $Tp'=Ts=\epsilon+p$. The term proportional to $(b{-}a')$ looks like a bulk viscosity contribution,  however it is an artefact of using the thermodynamic frame: the frame-invariant one-derivative scalar~\cite{Bhattacharya:2011tra} vanishes, and the $(b{-}a')$ term gives no bulk viscosity and no dissipation. Indeed this symmetry realization appears essentially Euclidean, and in equilibrium states gives Euclidean (zero frequency) correlators which analytically continue to retarded correlators in real time. 

In order to describe dissipative hydrodynamics, we need to add the $a$-type sources in the right way and some dynamical fields to the theory, in order to produce correlation functions which are not polynomial in spatial momenta. As noted earlier, the bottom-up construction (\ref{eq:S-lin-A}) and (\ref{eq:S-lin}) suggests a specific realization of the doubled symmetry ${\cal D}_1 \times {\cal D}_2$ in the hydrodynamic effective theory. Specifically, we assume an a nonlinear realization with an explicit (linear) realization of the diagonal ${\cal D}_r$ analogous to the treatment above. The presence of non-dynamical metric sources implies that the global symmetries of the theory are in effect weakly gauged.

We can be agnostic about the precise mechanism via which this symmetry realization arises from the microscopic theory, and carry out a general parametrization of the low energy degrees of freedom. This is the approach we take in this section, making use of the natural derivative expansion in the hydrodynamic regime. 
See~\cite{Endlich:2012vt, Grozdanov:2013dba} for a discussion of coupling the hydrodynamic degrees of freedom to a different type of Goldstone modes.

\subsection{Charge current}

Before considering diffeomorphisms in detail, to gain some intuition for the framework outlined above we first consider the simpler case of $U(1)$ charge diffusion in flat space. The above picture suggests that we should consider $U(1)_1\times U(1)_2 \rightarrow U(1)_r$ in the hydrodynamic regime. Denoting the corresponding `Goldstone-like' mode $\varphi_a$, on general grounds the low energy effective action will have the form
\be
 S_{\rm eff}[A_1^\mu,A_2^\mu,\varphi_a] = S_{\rm eff}[\xi_a^\mu,F_r^{\mu\nu}],
\ee
where $\xi_a^\mu = D^\mu \varphi_a = \ptl^\mu \varphi_a+ A_a^\mu$ is gauge invariant under the off-diagonal $U(1)_a$ symmetry, while $F_r^{\mu\nu} = \ptl^\mu A_r^{\nu} - \ptl^\nu A_r^{\mu}$ is gauge invariant under the residual $U(1)_r$. See Ref.~\cite{Son:2002zn} for a (technically) similar approach to the effective action for zero-temperature superfluids. The derivative expansion for the Goldstone modes is then equivalent to the expansion in the $a$-type fields. Note that consideration of any microscopic example of $U(1)_1 \times U(1)_2$ symmetry breaking provides some justification for the expectation that while $S_{\rm eff}[A_r,\xi_a]$ may naturally be represented in terms of hydrodynamic degrees of freedom at low energy, this need not be the case if the effective action is written in terms of the 1- and 2-type degrees of freedom.

To linear order in the $a$-fields we have
\be
\label{eq:S00}
 S_{\rm eff} = \int\!\! J_r^\mu (A^a_\mu + \ptl_\mu \varphi^a) + {\cal O}(a^2),
\ee
where $J_r^\mu = {\de S_{\rm eff}}/{\de A^a_\mu}$. The equation of motion for $\varphi^a$ is then $\ptl_\mu J^\mu_r=0$ as required. In writing~(\ref{eq:S00}), we have dropped the ${\cal O}(a^0)$ terms which only depend on the $r$-type fields and the $r$-type sources of the effective theory. These terms must ensure that the normalization of the generating functional $Z[A^r, A^a] =1$ for $A^a_\mu = \partial_\mu\lambda$ with $\lambda(t_0)=0$ is preserved in the effective description.

To determine the hydrodynamic constitutive relation for $J_r^\mu$, we write down a generic effective action depending on the allowed variables in the $r$ and $a$ sectors. We choose the variables for constructing the effective action to be given by the set
$
 \{T, u^\mu, \mu, \xi_a^\mu, F^{\mu\nu}_r\},
$
where for the moment we assume a flat background geometry, $T$ denotes a scalar that we identify with temperature, and $u^\mu$ is a normalized fluid velocity vector satisfying $u^\mu u_\mu = -1$. We can identify the timelike vector $\beta^\mu = \bar \beta^\mu + \beta^{\prime \mu}$ discussed above as $\beta^\mu = u^\mu/T$. In equilibrium, it is natural to instead start with $\beta^\mu$, which is then necessarily proportional to a timelike Killing vector, and construct the fluid velocity $u^\mu = \beta^\mu/\sqrt{-\beta^2}$ and temperature $T=1/\sqrt{-\beta^2}$. This implies a particular dependence on the background metric as a source, which we will not need below. We choose to start with the conventional hydrodynamic variables $T$ and $u^\mu$ for the following out-of-equilibrium analysis.

In terms of this data, up to ${\cal O}(a)$ there are three gauge invariant scalars at zeroth order in derivatives:
\begin{equation}
\label{eq:alphai}
   \alpha_1 = T\,,\ \ \ \alpha_2 = \mu\,,\ \ \ \alpha_3=\mu_a \,,
\end{equation}
where $\mu_a = u^\mu \xi_{a\,\mu}$. The identification of the $a$-sector chemical potential $\mu_a$ with the invariant $u^\mu \xi_{a\,\mu}$ (sometimes referred to as the Josephson relation) reflects a redundancy in the set of invariants, as both determine the source dual to the conserved charge,  see e.g. \cite{Herzog:2011ec}. We will adopt the identification above to remove this redundancy.
Note that $\al_{1}, \al_2 \sim {\cal O}(a^0)$, and $\al_3 \sim {\cal O}(a)$. Thus we find at this order,
\begin{align}
\label{eq:SD0}
 S_{\rm eff}^{(0)} = \int\!\!  F(T,\mu_r,\mu_a) + {\cal O}(a^2) \;
 = \int  F_{,\mu_a} u^\mu\;  D_\mu \varphi^a +{\cal O}(a^2),
\end{align}
where $F$ is the effective Lagrangian, and the derivative $F_{,\mu_a}\equiv \partial F/\partial\mu_a$ is evaluated at $\mu_a{=}0$. Comparing with (\ref{eq:S00}), the current is  $J_r^\mu = F_{,\mu_a} u^\mu + {\cal O}(a)$. Thus we can identify the charge density at this order as $n = \frac{\ptl p}{\ptl \mu_r} = F_{,\mu_a}$. 

At first order in derivatives and up to ${\cal O}(a)$, we have the following invariants:
$$
  \{\dot{\al}_i,\ \al_i',\ \partial_\mu u^\mu,\ \partial_\mu \xi_a^\mu,\  \dot u^\mu \xi_\mu^a ,\ u^\mu \dot \xi_\mu^a \,,\ u^\mu \xi^\nu_a F_{\mu\nu}^r \}.
$$
where $\dot{\al}_i = u^\mu \ptl_\mu \al_i$ and $\al'_i= \xi^\mu_a \ptl_\mu \al_i$. These terms can appear in the effective action, multiplied by coefficients which are functions of~$\alpha_i$. The number of such terms can be reduced by integrating by parts and redefining the relevant coefficients, and the action can be written as
\begin{equation}
\label{eq:SD1}
  S_{\rm eff}^{(1)} = \int \Big( c_1 \dot T + c_2 \dot \mu + c_3 \dot \mu_a + d_1 T' +  d_2 \mu' + d_3 \xi^\mu_a \dot u_\mu + d_4 \xi^\mu_a E_\mu  \Big) + {\cal O}(a^2)\,,
\end{equation}
where $c_i$ and $d_i$ are functions of $\alpha_i$, and $E_\mu = F_{\mu\nu}^r u^\nu$ is the $r$-type (physical) electric field. This again has the form of (\ref{eq:S00}). Combining the contributions from (\ref{eq:SD0}) and (\ref{eq:SD1}), the current to order ${\cal O}(a^0)$ is
\be
 J^\mu_r = {\cal N} u^\mu +j^\mu \,, 
 \ee
with $u^\mu j_\mu=0$, and
\begin{subequations}
\label{eq:Nj}
\begin{align}
  {\cal N} & = F_{,\mu_a} + (c_{1,\mu_a} {-} c_{3,T} {-} d_1)\dot T + (c_{2,\mu_a} {-} c_{3,\mu} {-} d_2) \dot\mu - c_3 \partial{\cdot}u\,,\\[5pt]
  j^\mu & = d_1 \Delta^{\mu\nu} \partial_\nu T + d_2 \Delta^{\mu\nu}\partial_\nu \mu + d_3 \Delta^{\mu\nu} \dot u_\nu + d_4 E^\mu\,.
\end{align}
\end{subequations}
It is reassuring to see the dissipative contributions to the current emerge from the effective action formulation. The constitutive relations (\ref{eq:Nj}) are written in a general ``frame'', and without using the ideal hydrodynamics equations of motion. The identification of the transport coefficients in terms of the parameters of the effective action needs to be done after the same expansion in the $a$-type fields is implemented for the energy-momentum tensor.

\subsection{Neutral fluid}

We now proceed along similar lines in considering the following realization of diffeomorphism invariance, ${\cal D}_1 \times {\cal D}_2 \rightarrow {\cal D}_r$ in the low energy hydrodynamic regime.
Linearized diffeomorphism invariance then ensures that the low energy effective action $S_{\rm eff}$ depends only on the combination $\xi_{\mu\nu}^a =  2{\cal D}_{\!\mu} \varphi^a_\nu = g_{\mu\nu}^a - \nabla_\mu \varphi_\nu^a - \nabla_\nu \varphi_\mu^a$ where $\varphi^a_\mu = g_{\mu\nu}^r \varphi_a^\nu$ is the vector `Goldstone-like' mode which transforms by a shift under $a$-type diffeos, and as a vector under $r$-type diffeos. Again,~$\nabla_\mu$ stands for $\nabla_\mu^r$, and the indices can be raised and lowered with the $r$-type metric. The effective action is then
\be
S_{\rm eff}[g_{\mu\nu}^1, g_{\mu\nu}^2, \varphi^\mu_a]=S_{\rm eff}[g_{\mu\nu}^r,\xi^a_{\mu\nu}]. 
\ee
It follows that 
\begin{equation}
\label{eq:STT}
 S_{\rm eff} = \int \coeff{1}{2} \sqrt{-g^r}\, T_r^{\mu\nu} (g^a_{\mu\nu} - \nabla_\mu \varphi^a_\nu - \nabla_\nu \varphi^a_\mu) + {\cal O}(a^2),
\end{equation}
where $\frac12\sqrt{-g^r}\, T^{\mu\nu}_r = {\delta S_{\rm eff}}/{\delta g^a_{\mu\nu}}$. The equation of motion for $\varphi^a_\mu$ is then $\nabla_\mu T_r^{\mu\nu}=0$ as required by (\ref{eq:Tr-conserv}) to ${\cal O}(a^2)$. Again, the unwritten ${\cal O}(a^0)$ terms must ensure the proper normalization of the generating functional in the effective theory.  The data available for constructing the effective action is given by the set
$
 \{T, u^\mu, g^r_{\mu\nu},\xi^a_{\mu\nu}\}.
$
As above, in equilibrium it is more natural to determine both the normalized fluid velocity $u^\mu = \beta^\mu/\sqrt{-\beta^2}$ and the temperature $T=1/\sqrt{-\beta^2}$ (where $\beta^2= g^r_{\mu\nu}\beta^\mu \beta^\nu$), in terms of a timelike vector $\beta^\mu$. We will not need to make assumptions about the metric dependence in the $r$-sector, so we will work with the conventional variables $T$ and $u^\mu$.  

In terms of this data, there are three scalars at zeroth order in derivatives up to ${\cal O}(a)$,
$$
  \alpha_1 =T\,,\ \ \ \ \alpha_4 = \xi_u\,,\ \ \ \ \alpha_5 = \xi_g\,,
$$
where $\xi_u \equiv u^\mu u^\nu \xi^a_{\mu\nu}$, and $\xi_g \equiv g_r^{\mu\nu} \xi^a_{\mu\nu}$.
Thus at zeroth order in derivatives
\begin{align}
\label{eq:STT0}
 S_{\rm eff}^{(0)} = \int \!\!\sqrt{-g_r}\, F(T,\xi_u,\xi_g) +{\cal O}(a^2)
  = \int \!\!\sqrt{-g_r}\, \left[F_{,\xi_u} u^\mu u^\nu + F_{,\xi_g} g_r^{\mu\nu} \right] \xi^a_{\mu\nu} +{\cal O}(a^2),
\end{align}
where $F$ is the effective Lagrangian, and the derivatives $F_{,\xi_u}\equiv \partial F/\partial\xi_u$ and $F_{,\xi_g}\equiv \partial F/\partial\xi_g$ are evaluated at $\xi_u=\xi_g=0$.
We can then read off $T^{\mu\nu}_r =  2F_{,\xi_u} u^\mu u^\nu + 2F_{,\xi_g} g_r^{\mu\nu} + {\cal O}(a)$, or
\be
 T^{\mu\nu}_r = (\epsilon+p)u^\mu u^\nu + p g_r^{\mu\nu} + {\cal O}(a)
\ee
on defining $\epsilon(T) \equiv 2F_{,\xi_u}-2F_{,\xi_g}$ and $p(T)\equiv 2F_{,\xi_g}$. Thus at zeroth order in derivatives the effective action gives rise to the energy-momentum tensor in ideal hydrodynamics.

At first order in derivatives and up to ${\cal O}(a)$, we have the following invariants:
$$
  \left\{ 
  \dot{\al}_i\,,  
  \nabla_{\!\mu} u^\mu\,, 
  u^\mu \xi_{\mu\nu}^a \partial^\nu T\,,
  \xi_{\mu\nu}^a \nabla^\mu u^\nu\,,
  u^\mu \xi_{\mu\nu}^a \dot u^\nu\,,
  u^\mu u^\nu \dot{\xi}^a_{\mu\nu}\,,
  u^\mu \nabla^\nu \xi_{\mu\nu}^a
  \right\}\,,
$$
where again the dot stands for $u^\mu \nabla_{\!\mu}$.
To order ${\cal O}(a)$, the coefficients of any structures involving derivatives of $\xi^a_{\mu\nu}$ can only depend on the invariant $T$, and integration by parts can be used to remove all terms with derivatives of $\xi_{\mu\nu}^a$ from the effective action, so that
\begin{equation}
\label{eq:STT1}
  S_{\rm eff}^{(1)} = \int\!\!\sqrt{-g_r} \left(
  f_1 \dot T + f_2 \nabla_{\!\mu} u^\mu + f_3  \xi_{\mu\nu}^a u^{(\mu} \nabla^{\nu)} T
  + f_4 \xi_{\mu\nu}^a \nabla^{(\mu} u^{\nu)} + f_5  \xi_{\mu\nu}^a u^{(\mu} \dot u^{\nu)}
  \right) + {\cal O}(a^2)\,.
\end{equation}
The brackets denote symmetrization, $\nabla^{(\mu} u^{\nu)} = \coeff12 (\nabla^{\mu} u^{\nu} + \nabla^{\nu} u^{\mu})$ etc. To linear order in the $a$-fields, the coefficients $f_1, f_2$ are functions of $T$, $\xi_u$ and $\xi_g$, while $f_3, f_4, f_5$ are functions of $T$ only. Expanding $f_1, f_2$ to first order in $\xi_{\mu\nu}^a$, we can read off the physical energy-momentum tensor from~(\ref{eq:STT}). Combining the contributions from (\ref{eq:STT0}) and (\ref{eq:STT1}), the result to order ${\cal O}(a^0)$ can be written as the standard hydrodynamic decomposition
\begin{equation}
\label{eq:TT-classical}
 T^{\mu\nu}_r = {\cal E} u^\mu u^\nu + {\cal P} \Delta^{\mu\nu} + (q^\mu u^\nu + q^\nu u^\mu) + t^{\mu\nu}\,,
\end{equation}
where $u_\mu q^\mu=0$, $u_\mu t^{\mu\nu}=0$, $g_{\mu\nu}^r t^{\mu\nu}=0$, and $\Delta^{\mu\nu}=g^{\mu\nu}_r + u^\mu u^\nu$. The coefficients are related to the parameters of the effective action by%
\begin{subequations}
\label{eq:const-rel-1}
\begin{align}
  {\cal E}  & = 2F_{,\xi_u}-2F_{,\xi_g} + 2(f_{1,\xi_u} {-} f_{1,\xi_g} {-} f_3)\dot T 
               +2(f_{2,\xi_u} {-} f_{2,\xi_g}) \nabla{\cdot}u\,,\\[5pt]
  {\cal P}  & = 2F_{,\xi_g} + 2f_{1,\xi_g}\dot T + 2(f_{2,\xi_g}{+}\coeff1d f_4) \nabla{\cdot}u\,,\\[5pt]
  q^\mu     & = f_3 \Delta^{\mu\nu}\partial_\nu T + (f_5 {-} f_4)\dot u^\mu\,,\\[5pt]
  t^{\mu\nu} & = f_4 \sigma^{\mu\nu}\,. 
\end{align}
\end{subequations}
Here $d$ is the number of spatial dimensions, and $\sigma^{\mu\nu} = \Delta^{\mu\alpha}\Delta^{\nu\beta}(\nabla_\alpha u_\beta + \nabla_{\!\beta} u_\alpha- \coeff2d\, g_{\alpha\beta}^r \nabla{\cdot}u)$ is the shear tensor. The derivatives with respect to $\xi_u$ and $\xi_g$ are evaluated at $\xi_u=\xi_g=0$. Expressions~(\ref{eq:const-rel-1}) should be viewed as constitutive relations in first-order hydrodynamics, obtained from the effective action for the hydrodynamic variables and Goldstone fields. The energy-momentum tensor~(\ref{eq:TT-classical}) is a classical ${\cal O}(a^0)$ quantity and will receive ${\cal O}(a)$ fluctuation corrections, as expected from~(\ref{eq:Tfluct}). The classical constitutive relations~(\ref{eq:const-rel-1}) are written in a general ``frame'', and without using the ideal hydrodynamics equations of motion. The ``frame'' is inherited from the effective action, similar to the thermodynamic frame of Ref.~\cite{Jensen:2012jh}. There are two transport coefficients in first-order classical hydrodynamics, the shear viscosity~$\eta$ and the bulk viscosity~$\zeta$. They can be identified from frame-invariant tensor ${\cal T}^{\mu\nu} = t^{\mu\nu} = - \et \si^{\mu\nu}$ and scalar ${\cal S} = {\cal P}^{(1)} - \frac{\ptl p}{\ptl \ep} {\cal E}^{(1)} = -\ze \nabla \cdot u$ combinations built from the one-derivative terms in~(\ref{eq:const-rel-1}) \cite{Bhattacharya:2011tra}. The identification of the bulk viscosity requires use of the zeroth-order scalar equations $u_\nu \ptl_\mu T^{\mu\nu}=0$ to relate the structures $\dot{T}$ and $\nabla \cdot u$. Making use of these relations, we find
$$
  \eta = -f_4\,,
$$
as well as
$$
  \zeta = -2(f_{2,\xi_g} {+} \coeff1d f_4) + 2\vs^2(T f_{1,\xi_g} {+} f_{2,\xi_u} {-} f_{2,\xi_g})
  -2\vs^4 T(f_{1,\xi_u} {-} f_{1,\xi_g} {-} f_3)\,,
$$
where $\vs^2 = \partial p/\partial\epsilon$ is the speed of sound squared.

\subsection{Charged fluid}
\label{sec-charged fluid}

We can generalize this discussion to the case of a charged fluid, by adding a background chemical potential. 
A priori, we can simply add the gauge field data for the $r$ and $a$ sectors to the metric data above. However, the additional vector $A_\mu^a$ transforms nontrivially under $a$-diffeos, see (\ref{eq:dAa}), and thus we need to modify the tensor data appropriately. We will work with the following basis of tensor data that has manifest invariance under $a$-gauge transformations and $a$-diffeos,
\be
 \{T, u^\mu, g^r_{\mu\nu}, \xi^a_{\mu\nu}, \mu, A^r_\mu, \ch^a_\mu\},
\ee
where $\ch^a_\mu = \xi^a_\mu - \varphi_a^\nu \ptl_\nu A^r_\mu - A^r_\nu \ptl_\mu \varphi_a^\nu$. (Note that $\nabla^r_{[\mu} \ch^a_{\nu]}$ depends only on $F_{\mu\nu}^r$ and is $a$- and $r$-gauge invariant.)  This data is manifestly invariant in the $a$-sector, and we can proceed to build $r$-gauge and $r$-diffeo invariants at the appropriate order in the derivative expansion. At ${\cal O}(a)$, the effective action is
\begin{equation}
\label{eq:STA}
  S_{\rm eff} = \int\!\! \coeff12 \sqrt{-g_r}\, T^{\mu\nu}_r\, \xi_{\mu\nu}^a
  +\int\!\! \sqrt{-g_r}\, J^\mu_r\, \chi_\mu^a + {\cal O}(a^2)\,,
\end{equation}
where as before $\frac12\sqrt{-g^r}\, T^{\mu\nu}_r = {\delta S_{\rm eff}}/{\delta g^a_{\mu\nu}}$, and $\sqrt{-g^r}\, J^{\mu}_r = {\delta S_{\rm eff}}/{\delta A^a_{\mu}}$.
The equations of motion for $\varphi_a$ and $\varphi^\mu_a$ are
\begin{align}
&\nabla_{\!\mu} J^\mu_r=0, \\
 &\nabla^\mu T^r_{\mu\nu} + A^r_\nu \nabla_{\!\rh} J_r^\rh = F_{\nu\mu}^r J^\mu_r \,,
\end{align}
respectively. The conservation of the $r$-current ensures $r$-gauge invariance as required.

There are five scalars at zeroth order in derivatives up to ${\cal O}(a)$,
\be
 \alpha_1=T\,,\ \ \ \  \alpha_2 = \mu\,,\ \ \ \  \alpha_3 = \mu_a\,,\ \ \ \  \alpha_4 = \xi_u \,,\ \ \ \  \alpha_5 = \xi_g \,,
\ee
where $\mu_a = u^\mu \ch^a_{\mu}$, $\xi_u = u^\mu u^\nu \xi^a_{\mu\nu}$, $\xi_g = g_r^{\mu\nu} \xi^a_{\mu\nu}$ as before. Note that $\mu_a$ is not manifestly $r$-gauge invariant, and we will need to ensure that current conservation in the $r$-sector imposes invariance in the final equations of motion. 
Thus at this order,
\begin{align}
 S_{\rm eff}^{(0)} &= \int \sqrt{-g_r}\, F(T,\xi_u,\xi_g,\mu,\mu_a) + \cdots 
 \label{eq:S0-charged}
\end{align}
which upon comparing with (\ref{eq:STA}) gives
$T^r_{\mu\nu} = 2F(T,\mu)_{,\xi_u} u^\mu u^\nu + 2F(T,\mu)_{,\xi_g} g_r^{\mu\nu} + {\cal O}(a)$, 
as well as $J_r^\mu = F(T,\mu)_{,\mu_a} u^\mu + {\cal O}(a)$.

At first order in derivatives and up to ${\cal O}(a)$, we can combine the analyses of the previous two subsections to find the following invariants: 
$$
  \left\{ 
  \nabla_\mu \chi_a^\mu, \chi^\mu_a \dot u_\mu, u^\mu \dot\chi_\mu^a, \chi^\nu_a E_{\nu}^r,
  \nabla_{\!\mu} u^\mu, 
  u^\mu \xi_{\mu\nu}^a \partial^\nu T,  u^\mu \xi_{\mu\nu}^a \partial^\nu \mu,
  \xi_{\mu\nu}^a \nabla^\mu u^\nu,
  u^\mu \xi_{\mu\nu}^a \dot u^\nu,
  u^\mu u^\nu \dot{\xi}^a_{\mu\nu},
  u^\mu \nabla^\nu \xi_{\mu\nu}^a
  \right\},
 $$
together with $\dot{\al}_i = u^\mu \ptl_\mu \al_i$ and $\al'_i= \chi^\mu_a \ptl_\mu \al_i$.
To order ${\cal O}(a)$, the coefficients of any structures involving derivatives of $\xi^a_{\mu\nu}$ and $\ch^a_\mu$ can only depend on the invariants $T$ and $\mu$, and integration by parts along with redefinitions of the other coefficients can be used to remove these terms from the effective action, so that
\begin{align}
\label{eq:ST2}
  S_{\rm eff}^{(1)} &= \int\!\!\sqrt{-g_r} \left( f_1 \dot{T} 
   + c_2 \dot{\mu} + c_3 \dot{\mu}_a + d_1 T' + d_2 \mu' + d_3 \ch^\mu_a \dot u_\mu + d_4 \ch^\mu_a E_\mu\right. \nonumber\\
  & \qquad\left.+ f_2 \nabla_{\!\mu} u^\mu + f_3  \xi_{\mu\nu}^a u^{(\mu} \ptl^{\nu)} T
  + f_4 \xi_{\mu\nu}^a \nabla^{(\mu} u^{\nu)} + f_5  \xi_{\mu\nu}^a u^{(\mu} \dot u^{\nu)} + f_6\xi_{\mu\nu}^a u^{(\mu} \ptl^{\nu)} \mu
  \right) + {\cal O}(a^2)\,.
\end{align}
 The notation for the coefficient functions has been chosen to match the earlier discussion as much as possible. To linear order in the $a$-fields, the coefficients $f_1, f_2, c_2$ are functions of $\al_i$, for $i=1,\ldots,5$, while $f_3, f_4, f_5$ are functions of $T$ and $\mu$ only. Following the earlier discussion, we can expand all the monomials to ${\cal O}(a)$ and read off the energy momentum tensor and charge current as follows
\begin{align}
\label{eq:TT-classical-2}
 T^{\mu\nu}_r &= {\cal E} u^\mu u^\nu + {\cal P} \Delta^{\mu\nu} + (q^\mu u^\nu + q^\nu u^\mu) + t^{\mu\nu}\,,\\
 J^\mu_r &= {\cal N} u^\mu +j^\mu \,.
\end{align}
The coefficients are related to the parameters of the effective action, and for the energy momentum tensor are given by%
\begin{subequations}
\label{eq:const-rel-2}
\begin{align}
  {\cal E}  & = 2F_{,\xi_u}{-}2F_{,\xi_g}+2(f_{1,\xi_u} {-} f_{1,\xi_g} {-} f_3)\dot T + 2(c_{2,\xi_u}{-}c_{2,\xi_g}-f_6)\dot\mu
               +2(f_{2,\xi_u} {-} f_{2,\xi_g}) \nabla{\cdot}u\,,\\[5pt]
  {\cal P}  & = 2F_{,\xi_g}+2f_{1,\xi_g}\dot T + 2c_{2,\xi_g} \dot\mu + 2(f_{2,\xi_g}{+}\coeff1d f_4) \nabla{\cdot}u\,,\\[5pt]
  q^\mu     & = f_3 \Delta^{\mu\nu}\partial_\nu T + f_6 \Delta^{\mu\nu}\partial_\nu \mu+ (f_5 {-} f_4)\dot u^\mu\,,\\[5pt]
  t^{\mu\nu} & = f_4 \sigma^{\mu\nu}\,,
  \end{align}
while the current is 
  \begin{align}
 {\cal N} & = F_{,\mu_a} + (f_{1,\mu_a} {-} c_{3,T} {-} d_1)\dot T + (c_{2,\mu_a} {-} c_{3,\mu} {-} d_2) \dot\mu + (f_{2,\mu_a} - c_3) \nabla{\cdot}u\,,\\[5pt]
  j^\mu & = d_1 \Delta^{\mu\nu} \partial_\nu T + d_2 \Delta^{\mu\nu}\partial_\nu \mu + d_3 \Delta^{\mu\nu} \dot u_\nu + d_4 E^\mu\,.
\end{align}
\end{subequations}
These constitutive relations are presented in a specific hydrodynamic frame, and it is useful to determine the frame-invariant transport coefficients. For the charged fluid, there is one tensor, one vector, and one scalar invariant at first order in the expansion \cite{Bhattacharya:2011tra}. These are usually identified with the shear viscosity~$\eta$, the conductivity~$\sigma$ and the bulk viscosity~$\zeta$. In terms of the coefficients in the effective action, the tensor invariant remains as for the neutral fluid ${\cal T}^{\mu\nu} = t^{\mu\nu} = - \eta \si^{\mu\nu}$, and we find again that
\be
  \eta = -f_4\,,
\ee
except that $f_4$ is now a function of both $T$ and $\mu$. 

The scalar invariant for the charged fluid takes the form ${\cal S} = {\cal P}^{(1)} - \frac{\ptl p}{\ptl \ep} {\cal E}^{(1)} - \frac{\ptl p}{\ptl n} {\cal N}^{(1)}$ in terms of the one-derivative data in (\ref{eq:const-rel-2}). This invariant depends on the three tensor structures $\dot{T}, \dot{\mu}, \nabla \cdot u$, but using the two longitudinal ideal hydrodynamic equations $u_\nu \nabla_\mu T^{\mu\nu}=F^{\nu\rh} u_\nu J_\rh = 0$ and $\nabla_\mu J^\mu=0$, we can write ${\cal S}= - \zeta \nabla \cdot u$, with the coefficient uniquely identified with the bulk viscosity. The result is lengthy, so we will not present it explicitly.

A new feature of the charged fluid is the existence of a vector invariant. In terms of the transverse structures in (\ref{eq:const-rel-2}), it is given by
${\cal V}^\mu = j^\mu - \frac{n}{\ep+p} q^\mu$, and depends on the four transverse vectors $\De^{\al\beta}\ptl_\beta T, \De^{\al\beta}\ptl_\beta \mu, \dot{u}^\al, E^\al$. The transverse ideal hydrodynamic equations impose only one constraint among these structures. To isolate the charge conductivity as the unique transport coefficient in this sector, we require additional constraints that follow, for example, from the equilibrium generating functional \cite{Banerjee:2012iz, Jensen:2012jh} (or from positivity of the local entropy production). Namely, there is in fact only one linear combination of these structures which is consistent with the background equilibrium state, $\De^{\al\beta} \ptl_\beta (\mu/T) - E^\al$. However, these additional constraints are not apparent in the effective action above, which was derived purely on the basis of a specific $(r,a)$ symmetry realization.

\section{Discussion}
\label{sec:discussion}
The top-down construction of the hydrodynamic effective action in Section~4 reproduces several features of the generating function for linearized hydrodynamics in Section~2. However, there are also certain missing elements, e.g. as noted for charged fluids in the preceding subsection. We therefore conclude by listing several questions left open by the present analysis. 
\begin{itemize}
\item
{\it (1,2) Basis}: We wrote down the hydrodynamic effective action by demanding $r$- and $a$-diffeo and gauge invariance, although only the $r$-sector symmetry was linearly realized. While this is sufficient to reproduce the expected tensor structures in the classical constitutive relations at ${\cal O}(a)$, going to higher orders in the $a$-expansion requires implementing the invariance under 1- and 2-sector symmetry transformations, for example in order to reproduce the conservation laws~(\ref{eq:WI-ra}). Manifest $(1,2)$ gauge and diffeo invariance is not straightforward in the $(r,a)$ basis, while the classical hydrodynamic equations are not straightforward to represent in the $(1,2)$ basis. Understanding the translation seems important for tackling several of the open questions below.

\item
{\it Equilibrium constraints}: In classical hydrodynamics, the existence of an equilibrium state in the presence of sources \cite{Banerjee:2012iz, Jensen:2012jh}, or (in some cases equivalently) the positivity of local entropy production, leads to powerful constraints on the possible thermodynamic response and transport coefficients. It is not immediately clear how to think about the entropy current from the point of view of the hydrodynamic effective action we have described, and in turn the source-dependence of the $r$-sector fields in equilibrium is not manifest. Clarifying these features would, for example, allow the correct identification of the electrical conductivity through a constraint among the coefficients of the tensors $\De^{\al\beta}\ptl_\beta T, \De^{\al\beta}\ptl_\beta \mu$ and $E^\al$.

\item 
{\it Fluctuation-dissipation constraints}: In addition to the equilibrium constraints above, which require the coefficients of certain tensor structures to vanish, it is apparent from the bottom up construction in Section~2 that there are nontrivial relations between different orders in the $a$-expansion of the effective action (\ref{S1}). For example, the dissipative transport coefficients enter both ${\cal J}_r$ and ${\cal K}_r$, as required to satisfy the fluctuation-dissipation theorem. It is important to understand the origin of these relations, and to determine any connection to the vanishing conditions above for specific coefficients in the classical constitutive relations.

\item
{\it External fields}: The charged fluid analyzed in Section~\ref{sec-charged fluid} raises a question of how $a$-sector diffeomorphism invariance can be manifest in the presence of the explicit violation induced by the background electromagnetic field. The vector $\ch_a^\mu$ is not manifestly invariant under $r$-sector gauge transformations, and indeed it seems clear that it should enter only in the combination $\mu_a = u_\mu\ch_a^\mu$. This leaves open the question of how such terms should enter at quadratic order in the $a$-expansion, as is required to restore the fluctuation-dissipation relation.

\item
{\it Path integral measure}: The hydrodynamic effective actions such as (\ref{eq:S0-charged}) and (\ref{eq:ST2}) are meant to be used in the path integral with both $r$-type  and $a$-type dynamical fields. Going beyond ${\cal O}(a)$ in the effective action requires understanding what the integration measure is for the $r$-type (physical) variables. The microscopic definition implies that $Z[g_1,g_1]=1$, and in the low energy regime the nontrivial measure factor required to ensure this was (formally) introduced as the functional ${\cal I}_r$. For the linearized hydrodynamics of Section~\ref{sec:bottom-up} the issue does not arise, as the action is linear in the $r$-type fields. Knowledge of the correct measure is important to determine off-shell interactions of the hydrodynamic degrees of freedom.

\item 
{\it Galilean hydrodynamics}: The focus of this paper has been on relativistic hydrodynamics. It would be interesting to develop this approach for systems with Galilean invariance, taking advantage of the recent understanding of Newton-Cartan sources in non-relativistic fluids \cite{Son:2013rqa,Jensen:2014ama,Jensen:2014wha}. 

\end{itemize}

\noindent {\bf Note Added}: As this paper was being finalized, the paper \cite{Haehl:2015pja} appeared on the arXiv. The latter work includes a comprehensive classification of non-dissipative transport to all orders in the hydrodynamic expansion, and has some overlap with Section~4 of the present paper in its treatment of dissipative terms within a CTP-like formalism. It would be interesting to understand the relations between these two approaches in more detail.

\subsection*{Acknowledgments}
We thank K.~Jensen, S.~Jeon and L.~Yaffe for helpful conversations. This work was supported in part by NSERC of Canada.\\

\pagebreak
\appendix

\bigskip
\noindent{\bf{\large Appendices}}

\section{Shear and sound response functions}

To see that that the generating functional (\ref{eq:ZZ2}) does indeed reproduce the required response functions given in (\ref{eq:GG-shear-sound}), let us go to Fourier space and integrate out the $\varphi^a$ fields. This will give rise to an effective action which is non-local, but real. With the properly normalized measure, we have
$$
  \int\!\! D\varphi^a_k\; e^{\int\! T\bar w\, \varphi^a_i M_{ij} \varphi^a_j}\, e^{i\!\int\! \varphi^a_i F_i} =
  e^{\,\int\!\! \frac{1}{4T\bar w} F_i (M^{-1})_{ij} F_j}\,.
$$
For the case at hand, $M_{ij} = -\gamma_\eta(\k^2\delta_{ij} {-} k_i k_j) - \gamma_s k_i k_j$, hence
$$
  (M^{-1})_{ij} = 
  -\frac{1}{\gamma_\eta \k^2} \left( \delta_{ij} - \frac{k_i k_j}{\k^2} \right)
  -\frac{1}{\gamma_s \k^2} \frac{k_i k_j}{\k^2}\,.
$$
The generating functional now becomes
$$
  Z[h_r,h_a] = \int\!\! D\pi_i \, 
  \exp \Big[\!\int_{\omega,\k}\!
  \frac{1}{4T\bar w}\, F_i[\pi,h^r] (M^{-1})_{ij} F_j[\pi,h^r] +{\cal O}(h^a) \Big]\,,
$$
where
$$
 F_i[\pi,h^r] = S_{ij}\pi_j 
                -i\bar w \omega h_{0i}^r - \frac12 \bar w i k_i h_{00}^r\,.
$$
We have defined 
$$
  S_{ij} = \Delta_\eta \left( \delta_{ij} - \frac{k_i k_j}{\k^2} \right)
          +\Delta_s \frac{k_i k_j}{\k^2}\,,\ \ \ \ 
  (S^{-1})_{ij} = \frac{1}{\Delta_\eta} \left( \delta_{ij} - \frac{k_i k_j}{\k^2} \right)
          +\frac{1}{\Delta_s} \frac{k_i k_j}{\k^2}\,,
$$
where $\Delta_\eta = (-i\omega + \gamma_\eta \k^2)$, $\Delta_s = (-i\omega + i\vs^2 \frac{\k^2}{\omega} + \gamma_s \k^2)$. Setting the sources to zero, we can evaluate the $rr$ function $\langle \pi_i \pi_j \rangle$, which is the symmetrized function (half the anti-commutator). With the sources set to zero, we have 
$$
  \langle \pi_i \pi_j \dots \rangle = 
  \int\!\! D\pi_i\; \exp \left[\frac12 \int
  \pi^\dagger \frac{S^\dagger M^{-1} S}{2T\bar w} \pi
  \right]
  \pi_i \pi_j \dots
$$
The $rr$ two-point function is therefore
$
  \langle \pi_i \pi_j \rangle = -2T\bar w \left(S^{-1} M S^{\dagger\, -1}\right)_{ij}\,,
$
or more explicitly
\begin{equation}
\label{eq:GRpiipij2}
  \langle \pi_i \pi_j \rangle  = \frac{i}{2}G_{\pi_i \pi_j}^{rr} = 
   \frac{2T\bar w\, \gamma_\eta \k^2}{|\Delta_\eta|^2} 
   \left( \delta_{ij} - \frac{k_i k_j}{\k^2} \right)
  +\frac{2T\bar w\, \gamma_s \k^2}{|\Delta_s|^2} \frac{k_i k_j}{\k^2}\,.
\end{equation}
This agrees precisely with the hydrodynamic $rr$ function (\ref{eq:Grr2}) found from the $ra$ function~(\ref{eq:GRpiipij}). The $rr$ functions involving $\epsilon$ can be obtained from the energy conservation constraint,
$$
  G_{\epsilon \epsilon}^{rr} = 
  \frac{k_i k_j}{\omega^2} G_{\pi_i \pi_j}^{rr}\,,\ \ \ \ 
  G_{\epsilon \pi_i}^{rr} = G_{\pi_i \epsilon}^{rr} = 
  \frac{k_j}{\omega} G_{\pi_i \pi_j}^{rr}\,.
$$
Again, they agree with the hydrodynamic $rr$ functions (\ref{eq:Grr2}) found from~(\ref{eq:GRepii}), (\ref{eq:GRee}). The $rr$ functions can of course be obtained by varying the generating functional (\ref{eq:ZZ2}) with respect to the $a$-type sources,
$$
  G_{\pi_i \pi_j}^{\,rr} = 2i\, \frac{\delta^2 Z[h_r,h_a]}{\delta h_{0i}^a\, \delta h_{0j}^a}\,.
$$
As usual, the sources are set to zero after the variation.
In order to find the $ra$ and $ar$ functions, we need to vary with respect to one $r$-source and one $a$-source. Looking at the generating functional (\ref{eq:ZZ2}), this will bring down one factor of $\pi_i$ and one factor of $\partial\varphi^a_i$. Specifically, in our conventions we have 
\begin{subequations}
\label{eq:GGZZ}
\begin{eqnarray}
  G^{\,ra}_{\pi_i \pi_j} =  
  i\,\frac{\delta^2 Z[h_r, h_a]}{\delta h_{0i}^a\, \delta h_{0j}^r} = 
  i\bar w\, \langle \pi_i\, \partial_t \varphi^a_j \rangle - \bar p \delta_{ij} \,,\\[5pt]
  G^{\,ar}_{\pi_i \pi_j} =  
  i\,\frac{\delta^2 Z[h_r, h_a]}{\delta h_{0i}^r\, \delta h_{0j}^a} = 
  i\bar w\, \langle \partial_t \varphi^a_i\, \pi_j \rangle - \bar p \delta_{ij}\,.
\end{eqnarray}
\end{subequations}
Integrating out $\varphi^a_0$ and $\epsilon$ in the generating functional (\ref{eq:ZZ2}) and setting the sources to zero, we have the following correlation functions:
$$
  \langle \pi_i\, \varphi^a_j \dots \rangle = 
  \int\!\! D\pi_k D\varphi^a_l\;
  e^{i\!\int\! \varphi^a_i S_{ij}\pi_j} e^{\int\! T\bar w\, \varphi^a_i M_{ij} \varphi^a_j}\;
  \pi_i \, \varphi^a_j \dots \,,
$$
with $M_{ij}$ and $S_{ij}$ defined above. This can be schematically represented using the combined field $\lambda_a = (\pi_i,\varphi^a_k)$ as
$$
  \langle \lambda_c\, \lambda_d \dots \rangle = 
  \int\!\! D\lambda \;
  e^{-\frac12 \int\! \lambda_a K_{ab}\lambda_b}\;
  \lambda_c \, \lambda_d \dots \,,
$$
where the matrix $K_{ab}$ in the $\pi \varphi^a$ space is
$$
  K = 
  \begin{pmatrix}
  0 & -iS^\dagger \\
  -iS & -2T\bar w\, M
  \end{pmatrix},\ \ \ \ \ 
  K^{-1} = 
  \begin{pmatrix}
  -2T\bar w\, S^{-1} M S^{\dagger\,-1} & iS^{-1} \\
  iS^{\dagger\,-1} & 0
  \end{pmatrix}\,.
$$
The correlation functions are given by $\langle \lambda_a \lambda_b\rangle = (K^{-1})_{ab}$, so we have
\begin{eqnarray*}
  && \langle \pi_i \pi_j \rangle = -2T\bar w \left(S^{-1} M S^{\dagger\,-1}\right)_{ij}\,,\\[5pt]
  && \langle \pi_i \varphi^a_j\rangle = i \left( S^{-1}\right)_{ij}\,,\ \ \ \ 
     \langle \varphi^a_i \pi_j\rangle = i \left( S^{\dagger\, -1}\right)_{ij}\,,\\[5pt]
  && \langle \varphi^a_i \varphi^a_j \rangle = 0\,.
\end{eqnarray*}
The $rr$ function is precisely what we have just evaluated in (\ref{eq:GRpiipij2}) by integrating out the $\varphi^a_i$ field first. For the mixed functions, we have from (\ref{eq:GGZZ})
$$
  G^{\,ra}_{\pi_i \pi_j} = 
  \left(\delta_{ij} - \frac{k_i k_j}{\k^2}\right)
  \left[ \frac{\bar w \gamma_\eta \k^2}{i\omega-\gamma_\eta \k^2} + \bar \epsilon \right]
  +
  \frac{k_i k_j}{\k^2}\,
  \left[\frac{\bar w\, \omega^2}
  {\omega^2-\k^2\vs^2+i\omega\gamma_s \k^2}  - \bar p \right]\,,
$$
as well as 
$$
  G^{\,ar}_{\pi_i \pi_j} = 
  \left(\delta_{ij} - \frac{k_i k_j}{\k^2}\right)
  \left[ \frac{\bar w \gamma_\eta \k^2}{-i\omega-\gamma_\eta \k^2} + \bar \epsilon \right]
  +
  \frac{k_i k_j}{\k^2}\,
  \left[\frac{\bar w\, \omega^2}
  {\omega^2-\k^2\vs^2-i\omega\gamma_s \k^2}  - \bar p \right]\,.
$$
The mixed functions agree with the response functions (\ref{eq:GRpiipij}) obtained by varying the on-shell hydrodynamic equations of motion, including the contact terms. The $aa$ functions all vanish due to $\langle\varphi^a_i \varphi^a_j\rangle=0$.

For the response functions involving the energy density we have
$$
  G_{\epsilon \pi_i}^{\, ra} 
  = 2i\, \frac{\delta^2 Z[h_r,h_a]}{\delta h_{00}^a\, \delta h_{0i}^r}
  = i \bar w \langle \epsilon\, \partial_t \varphi^a_i\rangle\,,\ \ \ \ 
  G_{\pi_i \epsilon }^{\, ra} 
  = 2i\, \frac{\delta^2 Z[h_r,h_a]}{\delta h_{0i}^a\, \delta h_{00}^r}
  = -i\bar w \langle \pi_i\, \partial_j \varphi^a_j \rangle\,.
$$
The factor of $2$ is due to the coupling of $h_{00}$. These can be evaluated by using $\epsilon = k_l \pi_l/\omega$, which is imposed in our generating functional (\ref{eq:ZZ2}), and the answer agrees precisely with~(\ref{eq:GRepii}). Similarly,
$$
  G_{\epsilon \epsilon}^{\, ra} 
  = 4i\, \frac{\delta^2 Z[h_r,h_a]}{\delta h_{00}^a\, \delta h_{00}^r}
  = \frac{k_i}{\omega} G_{\pi_i \epsilon}^{\,ra} - \bar\epsilon\,,
$$
where again the factor of $4$ is due to the coupling of $h_{00}$. This agrees precisely with (\ref{eq:GRee}), including the contact term.

\bibliographystyle{utphys}
\bibliography{Seff}

\end{document}